\documentclass[12pt,preprint]{aastex}
\usepackage{natbib,amssymb,amsmath}
\newcommand{\trecs}{T-ReCS}
\newcommand{\twhya}{TW~Hya}
\newcommand{\um}   {$\mu$m}
\newcommand{\ratzkaprivp}{(T.~Ratzka 2011, private communication)}
\newcommand{\Mjup}{$M_\mathrm{J}$}

\newcommand{\hotRingRadius}{2.9}           % AU
\newcommand{\hotRingTemperature}{570}      % Kelvin
\newcommand{\hotRingfRatio}{1e-4}          % dimensionless
\newcommand{\arnoldWallRadius}{3.9}        % AU
\newcommand{\arnoldWallTemperature}{116}   % Kelvins
\newcommand{\arnoldWallOuterRadius}{4.6}   % AU - arnoldWallRadius * (1 + calvetfRatio)
\newcommand{\arnoldSigma}{0.79}            % compared to Calvet SD
\newcommand{\arnoldKuruczScale}{1.0}       % Kelvins
\newcommand{\arnoldfRatio}{0.17}           % dimensionless

\newcommand{\hotCompanionRadius}{3.5}      % AU

\newcommand{\calvetWallRadius}{3.3}        % AU
\newcommand{\calvetWallTemperature}{130}   % Kelvins
\newcommand{\calvetfRatio}{0.11}           % dimensionless
\newcommand{\calvetKuruczScale}{2.0}       % dimensionless

\newcommand{\ratzkaAtmRadius}{0.6}         % AU
\newcommand{\ratzkaAtmTemperature}{186}    % Kelvins
\newcommand{\ratzkaWallRadius}{0.6}        % AU
\newcommand{\ratzkaWallTemperature}{271}   % Kelvins
\newcommand{\ratzkaWallfRatio}{0.051}      % dimensionless
\newcommand{\ratzkaAtmSigma}{1.2}          % compared to Calvet SD
\newcommand{\ratzkaKuruczScale}{1.9}       % dimensionless

\newcommand{\sublimationRadius}{0.02}      % AU
\newcommand{\sublimationTemperature}{1364} % Kelvins

\newcommand{\hotCompanionOrbitalPeriod}{2860}         % days

\newcommand{\deltaThetaInBetweenEpochs}{91}          % degrees

\newcommand{\companionTempAtEquilibrium}{150}         % Kelvins - equilibrium temperature at R_companion

\newcommand{\luminosityCompanion}{3-6$\times10^{-4}$}   % Ldot

\newcommand{\ourPredictedFluxFractionL}{0.8-3.0}          % percent, at L band

\newcommand{\ourPredictedMagContrastL}{5.3-3.6}           % mag, at L band

\newcommand{\ourPredictedFluxFractionM}{2-7}          % percent, at M band

\newcommand{\ourPredictedSeparation}{65}              % mas

\shorttitle{NEW MID-IR OBSERVATIONS OF TW HYA}{}
\shortauthors{ARNOLD ET AL.}{}
\slugcomment{Draft Version, 8 March 2012}{}

\begin{document}

\title{New Spatially Resolved Mid-IR Observations of the Transitional Disk
\twhya\ and Tentative Evidence for a Self-Luminous Companion}

\author{Timothy J. Arnold, J.~A. Eisner}

\affil{Steward Observatory, 
University of Arizona,
933 N. Cherry Ave,
Tucson, AZ 85721-0065, USA;
tjarnold,jeisner@email.arizona.edu}

\author{J.~D. Monnier}

\affil{Astronomy Department,
University of Michigan,
941 Dennison Bldg,
Ann Arbor, MI 48109-1090, USA;
monnier@umich.edu}

\author{Peter Tuthill}

\affil{School of Physics,
University of Sydney,
NSW 2006, Australia;
p.tuthill@physics.usyd.edu.au}

\begin{abstract}

We present spatially resolved observations of the canonical transition disk
object \twhya\ at 8.74~\um, 11.66~\um, and 18.30~\um, obtained with the \trecs\
instrument on the Gemini telescope. These observations are a result of a novel
observing mode at Gemini that enables speckle imaging.  Using this technique,
we image our target with short enough exposure times to achieve diffraction
limited images. We use Fourier techniques to reduce our data, which allows
high-precision calibration of the instrumental point spread function. Our
observations span two epochs and we present evidence for temporal variability
at 11.66~\um\ in the disk of \twhya. We show that previous models of \twhya's
disk from the literature are incompatible with our observations, and construct
a model to explain the discrepancies. We detect marginal asymmetry in our data,
most significantly at the shortest wavelengths. To explain our data, we require
a model that includes an optically thin inner disk extending from
\sublimationRadius\ to \arnoldWallRadius~AU, an optically thick ring
representing the outer disk wall at \arnoldWallRadius~AU and extending to
\arnoldWallOuterRadius~AU, and a hotter-than-disk-equilibrium source of
emission located at $\sim$\hotCompanionRadius~AU\@.

\end{abstract}

\section{Introduction}

Current planet formation theories are informed largely from observations
of evolved planetary systems. Since the discovery of the first extrasolar
planets \citep{MayorQueloz95} and the subsequent explosion of the
field, theories of planet formation have grown and proliferated. Though
the analysis of evolved systems is useful, perhaps the most effective way
to test planetary formation theories is through the direct observation of
young systems in the midst of planetary formation.

Objects thought to be in this stage, so-called transitional disks
\citep{Strom89}, are intermediate between embedded pre-main-sequence stars and
evolved planetary systems.  These objects exhibit evidence of a low-density,
optically thin gap in the innermost regions of their primordial circumstellar
disks. The presence of this gap was originally inferred from the near- and
mid-infrared~(mid-IR) properties of the spectral energy distributions~(SEDs).
The SEDs of these objects exhibit a deficit of flux at wavelengths less than
about 10~\um\ compared to the SEDs of classical T~Tauri stars (CTTS), and a
flux excess resembling CTT SEDs at wavelengths of $\sim$20-60~\um. The deficit
presumably occurs because the matter---and hence flux---in the inner regions of
the disk is absent.  The excess is caused by the outer disk, extending from the
puffed up ``edge'' or ``wall'' at several AU to the extent of the disk at tens
or hundreds of AU\@.

Though their geometries were originally inferred from the properties of their
unresolved SEDs, transition disk objects have since been resolved at various
wavelengths. \citet{Thalmann10} analyze $H$- and $K$-band data (1.65~\um\ and
2.2~\um, respectively) and resolve the gap in the transition disk LkCa~15. They
confirm a disk with a $\sim$46~AU truncation radius, inside of which is a
largely evacuated gap. This was also confirmed by \citet{Andrews11b} with
imaging of LkCa~15 at 870~\um.  Several other transition disks have been
resolved at millimeter wavelengths
\citep{Hughes07,Isella08,Brown09,Isella09b,Isella10,Andrews11a}.

The processes driving the formation of these gaps is not yet fully understood,
though several hypotheses have been proposed (photoevaporation, grain growth,
or a stellar or planetary companion).  Many recent observations have been
interpreted as evidence of a companion, supporting the hypothesis that young
planets are important in the evolution and dissipation of the disk in these
transition disk objects.  \citet{KrausIreland11} present the detection of a
likely protoplanet located inside the known gap in LkCa~15. They use
non-redundant aperture masking interferometry at three epochs to discover a
faint companion located at $\sim$16-21~AU from the primary star.
\citet{Huelamo11} detect a source in the $L$ band (3.5~\um) at a separation of
6.7~AU---within the disk gap---from the transition disk T~Chamaeleontis.
\citet{Eisner09} present mid-IR observations of the transition disk SR~21. They
spatially resolve the dust emission around this object, and suggest that the
disk around SR~21 must be completely cleared within $\sim$10~AU\@. They propose
a disk with a large inner hole and a warm companion near the outer edge of the
cleared region. 
 
TW Hydrae (\twhya), a classical T~Tauri Star, is one of the closest transition
disks, located in the nearby \citep[$\sim$50~pc;][]{Mamajek05} \twhya\
Association. Due in part to its proximity and age
\citep[$\sim$10~Myr;][]{Webb99}, \twhya\ has become an intensely studied
object.  \citet{Calvet02} first proposed that \twhya\ had a developing gap
located at $\sim$4~astronomical units~(AU), inferred from various features of
the SED\@. More recently, spatially resolved observations of \twhya\ have
offered additional insight.  The outer radius of the disk has been resolved in
millimeter wavelengths and determined to be in the range of $70-140$~AU
\citep{Wilner00,Wilner03,Qi04,Isella09a}.  \citet{Hughes07} resolved the inner
edge of the gap in the circumstellar disk at 7~mm, corroborating
\citet{Calvet02}'s determination of a hole at $\sim$4~AU\@.  Published
contemporaneously with \citet{Hughes07}, \citet{Ratzka07} contradicted the
claims of \citet{Calvet02} in their analysis of interferometric observations of
\twhya\ in the mid-IR with the Very Large Telescope Interferometer~(VLTI). They
suggested that the disk gap occurs at a considerably smaller radius:
$0.5-0.8$~AU\@. At yet shorter wavelengths, \citet{Eisner06} measured the
$K$-band~(2~\um) visibility at a single baseline with the Keck
Interferometer~(KI), inferring an inner radius for the optically thin,
evacuated region of $0.06$~AU\@.

Most recently, \citet{Akeson11} presented near-infrared~(near-IR)
measurements from the Center for High Angular Resolution Astronomy~(CHARA)
array and the KI at various baselines. They model many of the past
observations of \twhya\ simultaneously with theirs.  In combining the SED
and spatially resolved observations from the literature
\citep{Eisner06,Hughes07,Ratzka07}, they review and extend past attempts
at modeling the circumstellar disk, producing their own hybrid disk
geometry.  The \citet{Akeson11} model is essentially the original
\citet{Calvet02} model with an added optically thick ring of emission at
$\sim$0.5~AU, at roughly the disk equilibrium temperature (see
Section~\ref{sec:Modeling}).

Modeling of spatially unresolved data for \twhya\ (e.g. SEDs or higher
resolution spectra) can yield very different conclusions.  Spatially
resolved observations are necessary to properly constrain the disk
geometry.  It is important to note the ambiguity associated with modeling
source geometries from the SED alone: \citet[e.g.,][]{Boss93,Boss96} found
that the interpretation of infrared~(IR) flux deficits as central gaps
does not offer a unique solution, and that opacity and geometrical effects
produce degenerate solutions fitting the SED of an unresolved system. We
too find that this is true, and that the choice of dust species and thus
inner disk opacity can have a significant effect on the determination of
disk geometry.  Spatially resolved observations, ideally at multiple
wavelengths, are needed to unambiguously constrain disk geometry.

In order to further constrain models of \twhya's circumstellar disk, and
to resolve discrepant interpretations of observations, we present new
speckle interferometric observations in the mid-IR\@.  We observe \twhya\ at
three wavelengths (8.74~\um, 11.66~\um, and 18.30~\um) and at two epochs
(2007, 2009) using the Thermal-Region Camera Spectrograph (\trecs)
instrument on the Gemini telescope. We resolve the disk at each
wavelength, and construct a simple model to fit our observations.
Combining our results with previous spatially resolved imaging and
unresolved spectro-photometry, we constrain the properties of the inner
disk (or lack thereof) in \twhya.

Drawing from the literature, we first attempt to reproduce simple versions of
the models first proposed by \citet{Calvet02,Uchida04}, and \citet{Ratzka07}.
In the literature we identify two general classes of models: ``Calvet-like''
models with an optically thick wall at $\sim$4~AU with a hole of optically thin
material within \citep{Calvet02,Uchida04,Hughes07,Akeson11,Gorti11}; and
``Ratzka-like'' models for which there exists a similar opacity hole, but
located much closer in at $\lesssim$1~AU \citep{Ratzka07,Akeson11}. We
approximately reproduce both model types, and compute synthetic mid-IR
visibilities for comparison with our data; these are presented in
Section~\ref{sec:ExistingModels}. In Section~\ref{sec:NewDiskModel}, we
describe a new model consistent with both the data from the literature and our
new observations.

\section{Data}
\label{sec:Data}

\subsection{Observations}
\label{sec:Observations}

We present four observations of the transitional disk object \twhya\ at three
different wavelengths and two different epochs. A summary of these observations
is presented in Table~\ref{tab:Observations}. The wavelengths and epochs of the
observations are: 8.74~\um\ (8.74) and 11.66~\um\ (11.66a), obtained in 2007;
and 11.66~\um\ (11.66b) and 18.30~\um\ (18.30), obtained in 2009. For each
observation, we also took off-target nod pointings for infrared sky background
subtraction. Each nod, or pointing, consists of several short exposures
(``frames'') that are combined to yield a high resolution speckle image. It is
the individual frames, statistically combined, that are ultimately used in our
analysis. For each dataset, we observed a bright, point source calibrator in
order to calibrate and remove instrumental and atmospheric effects.

We collected our data using a custom observing mode on the \trecs\ instrument
at Gemini. We used short integration times ($t_\mathrm{int} \sim$172~msec) to
freeze the motion of the atmosphere and thus achieve high spatial resolution,
enabled by the diffraction limit of the telescope. We employed a dither / nod
pattern that moved observed objects around four positions on the detector. For
the 18.30~\um\ dataset, we employed standard chopping, and our frame
integration times were a factor of $\sim$10 longer than at the other
wavelengths. Total integration times are included in
Table~\ref{tab:Observations}.

The fluxes for our SED comparisons were obtained from the {\it Spitzer Space
Telescope} \/Infrared Spectrograph, from \citet{Uchida04} (see their Figure~2).
For our calculations of our best-fit model, we increase their stated errors of
$\sim$10\% to $\sim$30\% to account for the larger number of points in the SED
as compared to our resolved dataset, and to weight higher the value of the
resolved data over the unresolved data in our fits. The errors displayed in the
Figures are \citet{Uchida04}'s $\sim$10\%. We also use mid-IR interferometric
measurements for our analysis.  These data were obtained with the MIDI
instrument at the VLT, and were presented in \citet{Ratzka07}.

\begin{deluxetable}{rrrrrrrrrr}
\tabletypesize{\scriptsize}
\tablecolumns{10} 
\tablewidth{0pc} 
\tablecaption{Observations of \twhya}
\tablehead{
    \colhead{Date}                    &
    \colhead{Calibrator}              &
    \colhead{$\lambda$ (\um)}         &
    \colhead{$t_\mathrm{int}$ (msec)} &
    \colhead{$N_\mathrm{nod}$  }      &
    \colhead{$N_\mathrm{frm}$  }      &
    \colhead{$N_\mathrm{frm}^*$}      &
    \colhead{$T_\mathrm{int}$ (sec)} \\
    & & & & & & \\
    \colhead{(1)}  &
    \colhead{(2)}  &
    \colhead{(3)}  &
    \colhead{(4)}  &
    \colhead{(5)}  &
    \colhead{(6)}  &
    \colhead{(7)}  &
    \colhead{(8)}  }
\startdata
    9 May 2007   & II Hya    &  8.74 &  173  &  18 &  702 &  351 &  61  \\
    9 May 2007   & II Hya    & 11.66 &  173  &  30 & 1170 & 1131 & 195  \\
 9,18 April 2009 & HD 92036  & 11.66 &  181  & 100 & 3400 & 2505 & 454  \\
22,23 May 2009   & HD 92036  & 18.30 & 1813  &  76 &  228 &  228 & 413  \\
\enddata 
\tablecomments{The data used in our analysis. We observed the target, \twhya, at three
different wavelengths and at two different epochs. We discard unusable
frames, according to criteria in Section~\ref{sec:Reduction}. Columns are: 
(1) date of observation; 
(2) name of calibrator star; 
(3) wavelength of observation in microns;
(4) integration time of an individual frame in milliseconds (msec); 
(5) number of nods in the observation of the target; 
(6) number of individual frames in the observation of the target;
(7) number of frames used for analysis, after removing flagged frames; and
(8) total integration time for target using all usable frames, in seconds (sec).
}
\label{tab:Observations}
\end{deluxetable}

\subsection{Reduction and Analysis} 
\label{sec:Reduction}

\subsubsection{Fourier Analysis}

We reduce our data using Fourier analysis techniques. Each individual frame has
an exposure time that alone is too short to provide a significant signal to
noise ratio. However, a single long exposure would yield an image with
insufficient angular resolution due to the effects of the Earth's turbulent
atmosphere. One cannot na\"{i}vely add together the short, individual frames,
as the target centroid shifts on the sky due to atmospheric turbulence. We
instead combine the power spectra of individual frames.  This method of
addition is independent of translations of the image centroid position.
Furthermore, analyzing the data in Fourier space allows us to remove the
instrumental point spread function (PSF) with a point-source calibrator with
high accuracy. In Fourier space, we divide the power spectrum of the source by
the power spectrum of a point source calibrator.  This is considerably simpler
than deconvolution in the image plane. 

The first step in our data reduction procedure is to remove the sky background.
We obtained our data in such a way that each series of frames (short exposure
images) of a target is paired with a slightly offset series of frames of the
same target; we call these two sets ``adjacent nods''. This offset causes the
target to appear on two different regions of our detector.  We use these
adjacent nod pairs to overcome the high sky background in the mid-IR by
subtracting from each target its adjacent nod.  As each nod contains many
frames, the median value of all the frames in a particular nod is used for the
background subtraction.  The median value of an adjacent nod is subtracted from
each individual short-exposure frame. For example, in a single nod pair, the
median value for all the frames in Nod Type ``A'' is subtracted from each frame
in Nod Type ``B'', and likewise the median value of ``B'' frames is subtracted
from frames in ``A''.  This process is repeated for each pair of nods.

Next, we flag unusable frames, without actually modifying any of the data used
for analysis.  We first attempt to locate a bright point source in the median
image of a pointing: either the target---\twhya---or a calibrator. To locate
the position of the star in a pointing (or nod), we calculate the median image
for that nod, subtract from the median image the mean value of all the pixels,
and set negative pixel values to zero.  We identify ``hot pixels'' as pixels
with values more than three standard deviations larger than the mean pixel
value of the resultant image. These ``hot pixels'' are assigned a value equal
to the average value of their nearest neighbors using image convolution. This
technique removes isolated ``hot pixels'' while leaving signal from a star
relatively unaffected. The position of the star is then set to the location of
the maximally valued pixel in the image. We classify the pointing as unusable
if, after this process, there exist no pixels that are $>$5-$\sigma$ deviations
from the mean, or if the location of the point source determined by our
algorithm is closer than the width of our subimages to the detector edge (see
next paragraph). If we fail to detect a bright point source in a particular
pointing, we flag all the frames in that pointing as unusable. This could
happen due to poor seeing in these exposures, the presence of clouds, or some
other effect.  We note again that the steps described in this paragraph do not
ultimately affect the data used for analysis, but only for data flagging.
After a star is located, we return to the background-subtracted data described
at the end of the previous paragraph.

After eliminating flagged data, we are left with frames containing usable point
sources. For each frame, we cut out a ``postage-stamp'' size subimage from the
raw telescope image, centered on the point source. These subimages are
64x64~pixels, or 5.76$\times$5.76~arcseconds at \trecs's plate scale, or
$\sim$309$\times$309~AU at the distance of \twhya. We are then left with order
hundreds (ranging from $24$ to $3400$) of short-exposure frames for target and
calibrator for each dataset, cropped and centered at the star's position.  

For each frame, we subtract a residual sky value in addition to the initial IR
sky background subtraction: the median of pixel values in the outer regions
($48$ pixels or farther from the image center of each subimage). We then apply
a Hanning window to the image by multiplying by the two-dimensional Hanning
function: 
\begin{equation} 
    H(x,y) = 0.5\left(1+\cos\left(2 \pi
    \frac{\sqrt{(x-x_0)^2 + (y-y_0)^2}}{D}\right) \right)
\end{equation}
where $D$ is the size (diameter) and $(x_0,y_0)$ is the center of the Hanning
window. For our analysis, $D=16$~pixels (roughly $1.4$~arcseconds or $77$~AU).
The Hanning window has some useful properties: it is unity at the center and
falls smoothly to zero at the edges. This step removes spurious Fourier signals
by smoothing sharp brightness transitions. We take the Fourier transform of the
Hanning windowed image, and record the squared amplitude at each pixel (two
dimensional power spectrum, or power image).  We perform steps identical to
those described above for an adjacent, equally sized, background-subtracted
subimage of blank sky from the same parent image to produce a sky power image.

After computing the Fourier amplitudes, we use sigma-clipping to discard
discrepant data. For each dataset, we calculate the mean and standard deviation
of the squared visibility as a function of baseline for both target and
calibrator. If any individual frame deviates $>$3-$\sigma$ from the mean at any
moderate, well-behaved baseline (between 0~m and 5~m), we exclude it from the
sample.  This step excludes a small fraction of our resulting frames: usually
zero, but occasionally as high as 2$\%$.

\subsubsection{Calibration}
\label{sec:Calibration}

After the reduction procedure outlined above, we are left with several hundred
power images, for both target and calibrator. Additionally, we have power
images for adjacent, blank sections of sky from parent images for each of these
postage-stamp images.  We first remove any bias due to detector artifacts by
subtracting the sky power image from the object power image. We then subtract a
residual bias: any non-zero value for the visibility amplitude at baselines
well beyond our sensitivity (9~m and longer).  We set negative values of the
target's power images to zero, and negative values of the calibrator to a small
value ($10^{-5}$; to avoid eventual division-by-zero computational errors).  We
then normalize the target and calibrator power images by the maximum value in
each respective set. 

We calculate an azimuthal average of each power image by computing the mean of
the pixels in an annulus corresponding to a particular baseline length. The
power obtained in each annulus becomes the visibility amplitude at a
corresponding baseline. We do this to increase the signal to noise and to make
our plots more straight-forward to interpret.  Furthermore, to rough
approximation, it is an acceptable assumption that our target---a star with a
nearly face-on circumstellar disk---is indeed symmetric about the azimuth.
\citet{Calvet02} also assume a face-on orientation, in agreement with estimates
of \twhya's inclination in the literature \citep{Krist00,Qi04,Pontoppidan08}.
We discuss possible departures from our assumption of azimuthal symmetry in
Section~\ref{sec:DiskCompanionModel}.

We use the azimuthally averaged visibility curves (one for each frame) to
calculate a mean and standard deviation of the mean (SDOM, or standard error)
for target and calibrator. We then use these derived values ($V_\mathrm{trg}$,
$\sigma_\mathrm{trg}$, $V_\mathrm{cal}$, $\sigma_\mathrm{cal}$) to calculate
the calibrated visiblity ($V_\mathrm{trg}/V_\mathrm{cal}$) and its associated
statistical uncertainty. The SDOM is calculated by dividing the usual standard
deviation by the square root of the number of frames ($\sqrt N_\mathrm{frm}$).

The total statistical uncertainty associated with the calibrated visibility is
given by the propagation of the two SDOMs. We calculate the propagated error
using the usual relation, 
\begin{equation}
\sigma_f^2 = \left(\frac{\partial f}{\partial a}\right)^2 \sigma_a^2
           + \left(\frac{\partial f}{\partial b}\right)^2 \sigma_b^2 
\end{equation}
for $f = f(a,b)$ and for uncorrelated errors in $a$ and $b$. 

\begin{figure}
    \epsscale{0.7}
    \plotone{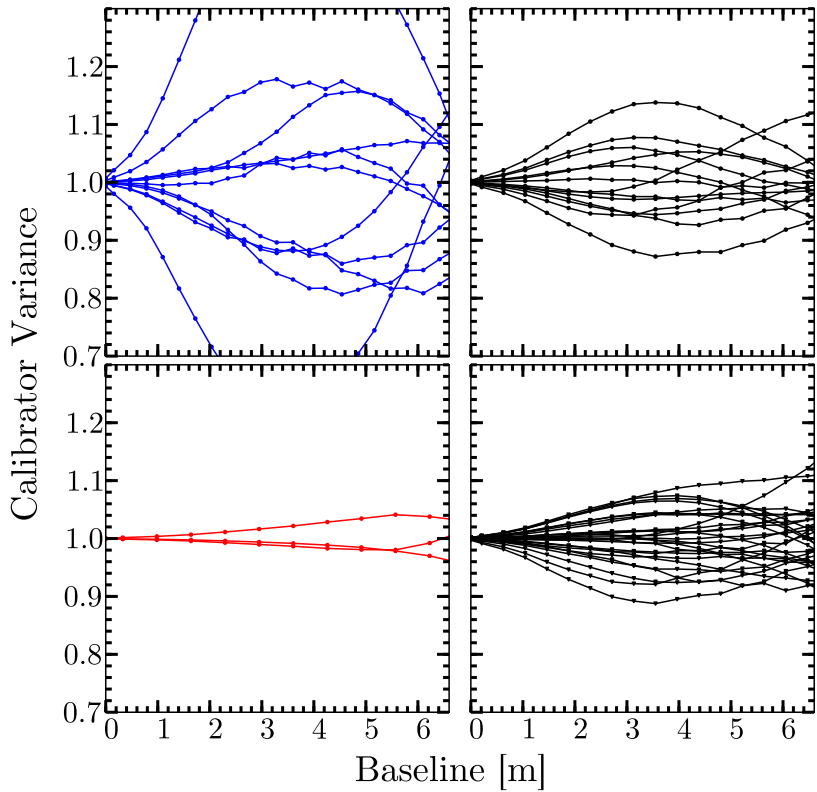}
    \caption{In this Figure, we illustrate the method used to estimate
systematic errors in our visibility measurements. For each observation
block, we plot the value of the visibility for a calibrator as a function
of baseline, divided by the mean value for all the blocks. We perform this
test for each dataset, shown in the separate panels: 8.74 ({\it blue, top
left}\/), 11.66a ({\it black, top right}\/), 11.66b ({\it black, bottom
right}\/), and 18.30 ({\it red, bottom left}\/). The value of the systematic
error estimate is the standard error (or standard deviation of the mean,
SDOM) of the observation blocks at each baseline.
    \label{fig:CalCal}}
\end{figure}

We also estimate a systematic error for our experimental setup, by comparing
calibrator observations. For example, variation in seeing or sky background
structure over timescales longer than our source-calibrator cadence will lead
to systematic errors. Similar to the procedure described above, we average the
azimuthal average of power images according to telescope pointing (i.e., we
group calibrator frames of a single pointing together). We divide
pointing-averaged calibrator visibility curves by the mean visibility curve of
all the calibrators. We use the SDOM---not the standard deviation---of the
several calibrator pointings as a systematic error: the magnitude at which we
expect an uncertainty to exist for a given set of observations. For each
baseline, the visibilities of the calibrators do appear to be normally
distributed. We add this error in quadrature with the statistical uncertainties
described above. The final error in our measurements is dominated by these
estimated systematic errors. We present the data used for this method of
systematic error estimation in Figure~\ref{fig:CalCal}. This plot shows the
visibility amplitude of each calibrator pointing, divided by the mean
visibility amplitude of all the calibrators, as a function of baseline, for
each dataset.

We expect the calibrator errors to appear correlated to some extent. That is,
if a calibrator has a larger visibility amplitude at one baseline, it likely
has a larger visibility at other baselines as well, as seeing variations are an
important cause of the dispersion. One other possible cause for correlated
errors is our application of a Hanning windown in the data reduction process.
To check this effect, we perform our analysis without the Hanning window; we
notice a slight decrease in apparent correlation. The size of the calculated
error bars did not change appreciably, however, and we keep the Hanning window
in our reduction process.

\subsection{Size of Emitting Region in \twhya}
\label{sec:Size}

\begin{figure}
    \epsscale{0.7}
    \plotone{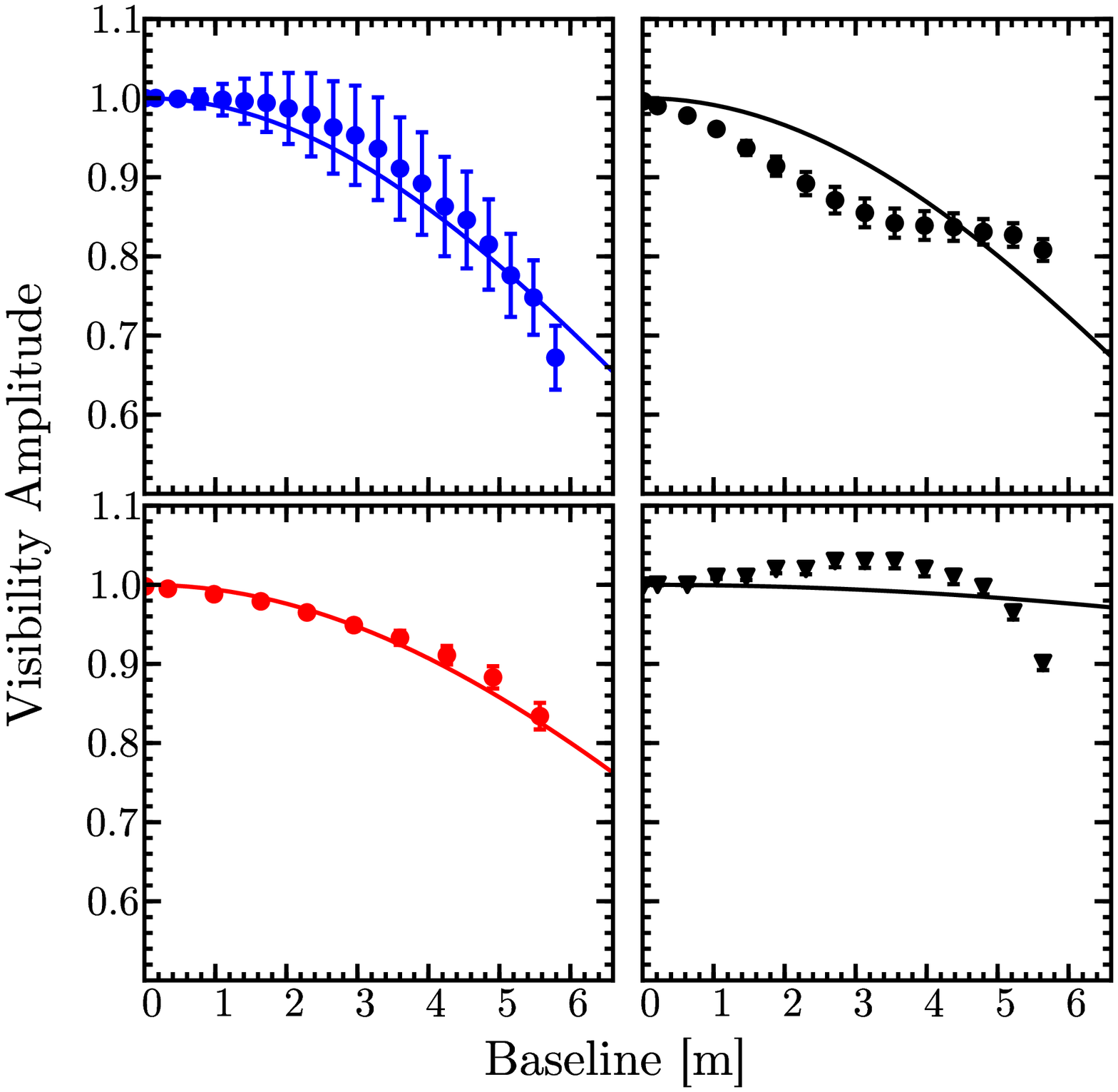}
    \caption{
This figure shows the visibilities predicted by ring models of various sizes.
The rings emit at a single wavelength and have width to radius ratio of $f =
\calvetfRatio$.  This exercise shows that our 8.74~\um\ observations ({\it
blue, top left}\/) are significantly more resolved than all of our other epochs
The four separate observations are: 8.74 ({\it blue, top left}\/), 11.66a ({\it
black, top right}\/), 11.66b ({\it black, bottom right}\/), and 18.30 ({\it
red, bottom left}\/). 
\label{fig:Size}}
\end{figure}

We estimate the size of the emitting region at each wavelength by fitting a
ring of fixed thickness (width to radius ratio, $f = $\calvetfRatio) to each
visibility dataset (see Equation~\ref{eq:VisibRing}). We performed a chi
squared minimization where the size of the ring of emission was the only free
parameter. Error bars (1-$\sigma$) were derived by finding the ring size value
that corresponded to the minimum chi squared plus one. The results of our fit
can be seen in Table~\ref{tab:Size} and a comparison of our data with the best
fit ring visiblity curves can be found in Figure~\ref{fig:Size}.

\begin{deluxetable}{crr} 
\tabletypesize{\scriptsize}
\tablecolumns{3} 
\tablewidth{0pc} 
\tablecaption{Single Wavelength Ring Fit Results} 
\tablehead{ 
    \colhead{Observation Name}     & 
    \colhead{$R_\textrm{in}$ [AU]} & 
    \colhead{$f$}                  \\
    & & \\
    \colhead{(1)}  & 
    \colhead{(2)}  & 
    \colhead{(3)}  }
% updated this table on 25-Oct 2011
\startdata 
     874   & $1.98\pm0.04$ & \calvetfRatio \\
     1166a & $2.56\pm0.23$ & \calvetfRatio \\
     1166b & $0.71\pm0.50$ & \calvetfRatio \\
     1830  & $3.35\pm0.31$ & \calvetfRatio \\
\enddata 
\tablecomments{
The results of a single wavelength ring fit to our observations. Error
bars are derived by finding the value (in radius) on the chi squared
contour that corresponds to one plus the minimum chi squared value.
Columns are: (1) name of observation; (2) best fit radius with
$1$-$\sigma$ error bars; (3) the ratio of ring width to inner radius, set
to a constant for these illustrative purposes. We note the
$\gtrsim$2.5-$\sigma$ difference in size between our two epochs of
11.66~\um\ observations.
}
\label{tab:Size}
\end{deluxetable} 

\subsubsection{Variability}
\label{sec:Variability}

Since we have two epochs at 11.66~\um, we are able to constrain
variability over a timescale of roughly two years ($\sim$720 days).
Variability is observed between these two epochs at the
$\gtrsim$2.5-$\sigma$ level (Figure~\ref{fig:Size}; Table~\ref{tab:Size}).
For simplicity, we use both epochs of 11.66~\um\ data in our modeling. We
speculate as to the cause of this variability in
Section~\ref{sec:Discussion}.

\section{Modeling}
\label{sec:Modeling}

We model the \twhya\ system using largely geometric models. We approximate the
central star as the Kurucz stellar atmosphere of a K7 star with radius $R_* =
R_\odot$, temperature $T_* = 4000$~K, surface gravity
log~$g$~(cm~s$^{-2}$)~=~4.5, at a distance $d = 53.7$~pc
\citep{Webb99,vanLeeuwen07}. We note, as has been mentioned previously in the
literature \citep[e.g.,][]{Sitko00}, that \twhya\ is a known variable star, so
we do not necessarily expect a Kurucz model to exactly model the emission of
the star. Also, the shorter-wave flux density measurements from the literature
were not taken contemporaneously with the mid-IR spectrum that comprises the
bulk of the disk emission. While this Kurucz model does not perfectly reproduce
the stellar flux at all wavelengths, and different analyses of \twhya\ use
slightly different stellar model parameters, it is sufficient for our modeling
purposes. 

Our models of \twhya\ are composed of an optically thin disk (approximated as a
series of concentric rings which follow a known temperature-radius
relationship) and an optically thick ring of a single temperature representing
the directly illuminated edge of the optically thick disk.  For the optically
thin component, we must choose and apply dust opacities.  We discuss the
details of this choice below.

\subsection{Existing Models}
\label{sec:ExistingModels}

We first attempt to reproduce the large cavity model of
\citet{Calvet02,Uchida04}. We use geometric model parameters and dust opacities
from their work. We extract dust optical depth directly from
\citet{Uchida04}\footnote{To obtain the optical depths used by
  \citet{Uchida04}, we first record the reported flux density from their
  optically thin disk component (see their Figure~2, top panel). We then use
  these published flux densities, an assumption of constant surface density (as
in \citet{Calvet02}), and their stated disk geometry, to extract values for the
dimensionless wavelength-dependent disk optical depth, $\tau_\lambda$, that
they use.}, which presents results consistent with those of \citet{Calvet02}.
This model consists of emission from three components: 1) a star, 2) an
optically thin disk, extending from the dust sublimation radius at
\sublimationRadius~AU to the directly illuminated disk wall at
\calvetWallRadius~AU, and 3) an optically thick ring at \calvetWallRadius~AU,
representing the puffed or flared wall of the outer disk edge.  To approximate
\citet{Calvet02,Uchida04}'s model of the stellar flux, we allow the flux
density produced by the Kurucz stellar model to vary so that it matches the
observed flux density at $4-5$~\um; this is similar to their own strategy and
the procedure of \citet{Sitko00}.

The flux from the optically thick disk wall is approximated by an unmodified
blackbody. The flux density of an annulus of thickness $dR$ is given by: 
\begin{equation} 
    dF_\lambda = \tau_\lambda \frac{2 \pi}{d^2} 
                 B_\lambda(T_\mathrm{dust}) R dR.
    \label{eq:dF} 
\end{equation} 
In the case of the optically thick ring, $\tau_\lambda = 1$ for all values of
$\lambda$. For the optically thin disk, the temperature profile is given by
\begin{equation} 
    T_\mathrm{disk} = 
    T_\mathrm{*} \left( \frac{R_\mathrm{*}}{2 R}\right)^{1/2},
    \label{eq:TOpticallyThin}
\end{equation}
where $T_\mathrm{*}$ is the temperature of the star, $R_\mathrm{*}$ is the
radius of the star, and $R$ is stellocentric radius.  To calculate the flux
from the disk, we use small concentric annuli. The temperature structure at
every point in the optically thin disk behaves according to
Equation~\ref{eq:TOpticallyThin}, and the fluxes from each annulus are summed
together to obtain the total disk flux. The temperature for the directly
illuminated rim or edge of the optically thick disk is given by
\begin{equation} 
    T_\mathrm{rim} = 
            T_\mathrm{*} \left( \frac{R_\mathrm{*}}{R}\right)^{1/2}.
    \label{eq:TOpticallyThick}
\end{equation}
We are motivated in the choice of these temperature profiles by the methods of
\citet{ChiangGoldreich97,Calvet02,Eisner06} (see Equation~1, Equations~1~and~3,
and Equation~1, respectively).

All model components can be viewed as collections of single-temperature rings
that obey the model temperature profile.  Model fluxes are computed by summing
Equation~\ref{eq:dF} over all annuli. The squared visibility for a ring (or
annulus) is given by \citep[see][]{Eisner03}:
\begin{equation}
 V_\mathrm{ring} = \frac{2}{\pi r_{uv} \theta_\mathrm{in} (2f + f^2)} \times
          \left[ (1+f) J_1\left[(1+f \pi \theta_\mathrm{in} r_{uv} \right] - 
          J_1(\pi \theta_\mathrm{in} r_{uv}) \right],
\label{eq:VisibRing}
\end{equation} 
where $r_{uv} = \sqrt{u^2 + v^2} = {\boldsymbol B} \cdot {\boldsymbol s} /
\lambda$ is the $u$-$v$ radius, $\lambda$ is the wavelength, $\theta$ is the
angular diameter of the object in radians, $f$ defines the ratio of radial
thickness to size of the annulus ($f = W / R$, where $W$ is the width of the
annulus and $R$ is the inner radius of the annulus), $J_1$ is the Bessel
function of the first kind of order one. The normalized visibility for the
entire model is the flux-weighted average of the visibilities produced by each
annulus. 

\begin{figure}
    \epsscale{0.7}
    \plotone{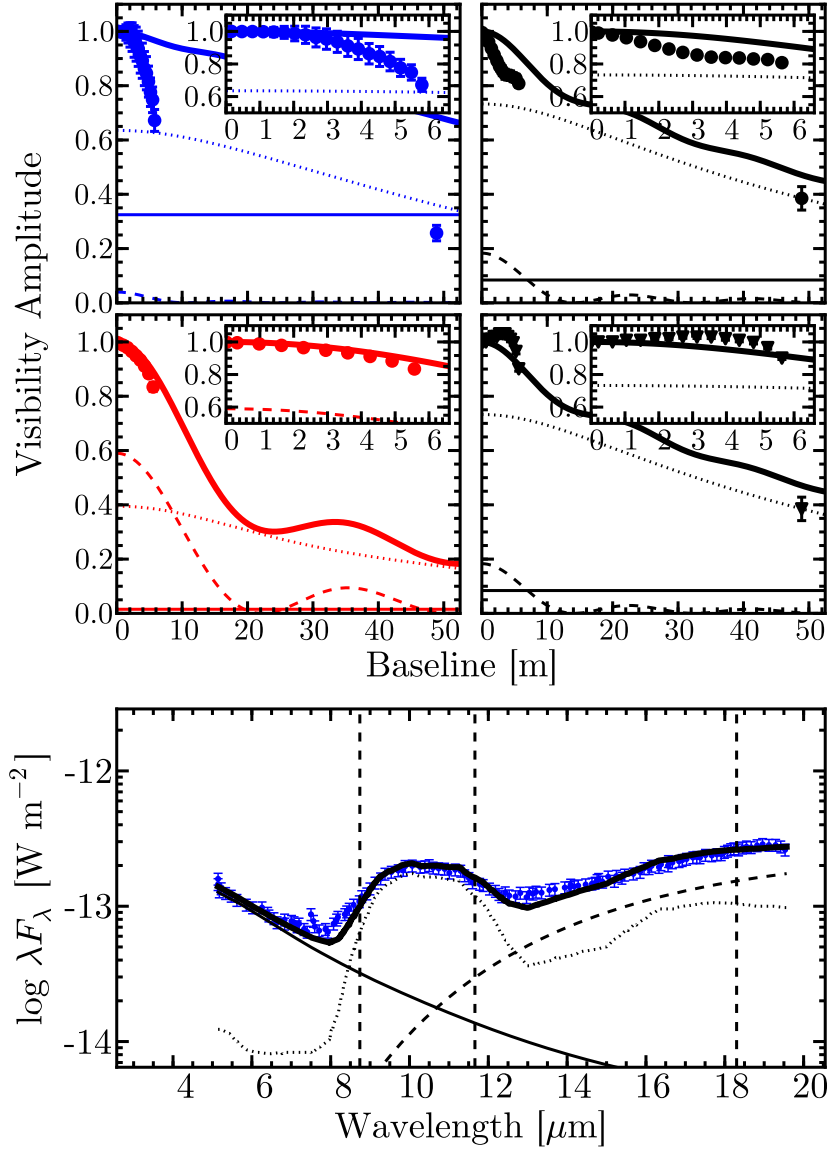}
    \caption{
The calculated SED and visibility curves for our realization of
\citet{Calvet02}'s model. See Table~\ref{tab:Models} for model parameters.
The large bottom panel shows the SED, indicating the flux from the star
({\it thin solid line}\/), the optically thin inner disk ({\it dotted
line}\/), the optically thick, directly illuminated disk wall ({\it dashed
line}\/), and the total ({\it thick solid line}\/). The flux density data
are also shown ({\it blue circles}\/). In the top half of the figure, we
show our four separate observations: 8.74 ({\it blue, top left}\/), 11.66a
({\it black, top right}\/), 11.66b ({\it black, bottom right}\/), and 18.30
({\it red, bottom left}\/). In each, the flux weighted visibilities of the
different model components are indicated by different line types, as in
the SED plot. The insets in the top right of each panel show the data
presented in this work, while the long baseline point at $\sim$45~m is
from \citet{Ratzka07}. Vertical, dashed lines in the SED panel show our
wavelengths of observation.
    \label{fig:Calvet}}
\end{figure}

From these model parameters (matched to those used in
\citet{Calvet02,Uchida04}), we generate a synthetic SED and visibilities
(Figure~\ref{fig:Calvet}).  The large bottom panel shows the SED, indicating
the stellar flux density ({\it thin solid line}\/), the optically thin, low
density inner disk ({\it dotted-dashed line}\/), the outer disk wall ({\it
dotted line}\/), and the total flux ({\it thick solid line}\/).  The flux data
are also shown ({\it blue circles}\/). In the top half of the figure, we show
our four separate observations: 8.74 ({\it blue, top left}\/), 11.66a ({\it
black, top right}\/), 11.66b ({\it black, bottom right}\/), and 18.30 ({\it
red, bottom left}\/). In each, the flux weighted visibilities of the different
model components are indicated by different line types, as in the SED plot. The
insets in the top right of each panel show zoom-ins of the data presented in
this work, while the long baseline points at $\sim$45~m are from
\citet{Ratzka07}.  Vertical, dashed lines in the SED panel show the wavelengths
of our observations at 8.74~\um, 11.66~\um, and 18.30~\um.

Our estimate of \citet{Calvet02}'s model reproduces the SED as well as done by
the original authors. This is not unexpected, as we obtained the opacities,
physical geometries, and temperatures directly from their work.  More
interesting are the mid-IR visibilities produced by this model, shown in the
top half of the figure. This model reproduces well the 18.30~\um\ data, a
tracer of cooler portions of the disk, dominated by emission of the transition
disk wall.  Similarly, this model reproduces rather well both epochs of our
11.66~\um\ data.  One epoch (1166b) is reproduced better than the other
(1166a), but we note that these two datasets are in fact inconsistent with each
other due to temporal variability (see Section~\ref{sec:Variability}). This
model is clearly inconsistent, however, with our very resolved 8.74~\um\
observations, as seen in the top left panel of Figure~\ref{fig:Calvet}. 

\subsubsection{Small Cavity Models}
\label{sec:SmallCavityModels}

Though subsequent work has confirmed the validity of \citet{Calvet02}'s model
\citep[e.g.,][]{Hughes07}, \citet{Ratzka07} claimed that the evacuated cavity
was much closer to the star, at $\lesssim$1~AU\@.  Since \citet{Ratzka07}'s
model was based largely on mid-IR flux measurements, and presented resolved
mid-IR data, we compare their work against our observations.  \citet{Ratzka07}
used a more complicated disk modeling technique than did \citet{Calvet02}; we
use our simple geometric models to distill out the most important elements. The
main sources of the mid-IR emission in the model of \citet{Ratzka07} were the
optically thick, directly illuminated disk wall, and its optically thin
atmosphere. We were thus motivated to produce emission at $\lesssim$1~AU from
nearly coincident components, one optically thick and one optically thin.

Our methods to reproduce the \citet{Ratzka07} model are similar to the process
described previously in Section~\ref{sec:ExistingModels}, with one exception.
As the bulk of the optically thin emission from the \citet{Ratzka07} model
originates from the disk atmosphere, we use an optically thin ring at a single
temperature ($R_\mathrm{atm}$, $T_\mathrm{atm}$) instead of the optically thin
inner disk described above ($R_\mathrm{in}$, $T_\mathrm{in}$). The temperature
of this ring is a model parameter and is free to vary. We also vary the surface
density of the optically thin ring, $\Sigma_\mathrm{atm}$.  We vary these free
parameters and minimize chi squared of the model compared to the mid-IR flux of
\twhya\ and the long-baseline ($\sim$45~m) visibility points from
\citet{Ratzka07}. As our motivation was to reproduce the model of
\citet{Ratzka07}, we do not include our new visibility data in the minimization
of chi squared.

\citet{Ratzka07} used different optical depths than did
\citet{Calvet02,Uchida04}, and we obtained and used these in our modeling
\ratzkaprivp. As in the modeling described above, we allow the flux density
produced by the Kurucz stellar model to vary so that it matches the observed
flux density at $4-5$~\um; this is similar to their own strategy and the
procedure of \citet{Sitko00}.

We performed a chi squared minimization and allowed the following parameters to
vary freely to generate a model consistent with the \citet{Ratzka07} geometry:
Kurucz scaling, $R_\mathrm{atm}$, $T_\mathrm{atm}$, $\Sigma_\mathrm{atm}$,
$R_\mathrm{wall}$, $f_\mathrm{wall}$, $T_\mathrm{wall}$.  These properties and
their best-fit values are listed in Table~\ref{tab:Models}.

\begin{figure}
    \epsscale{0.7}
    \plotone{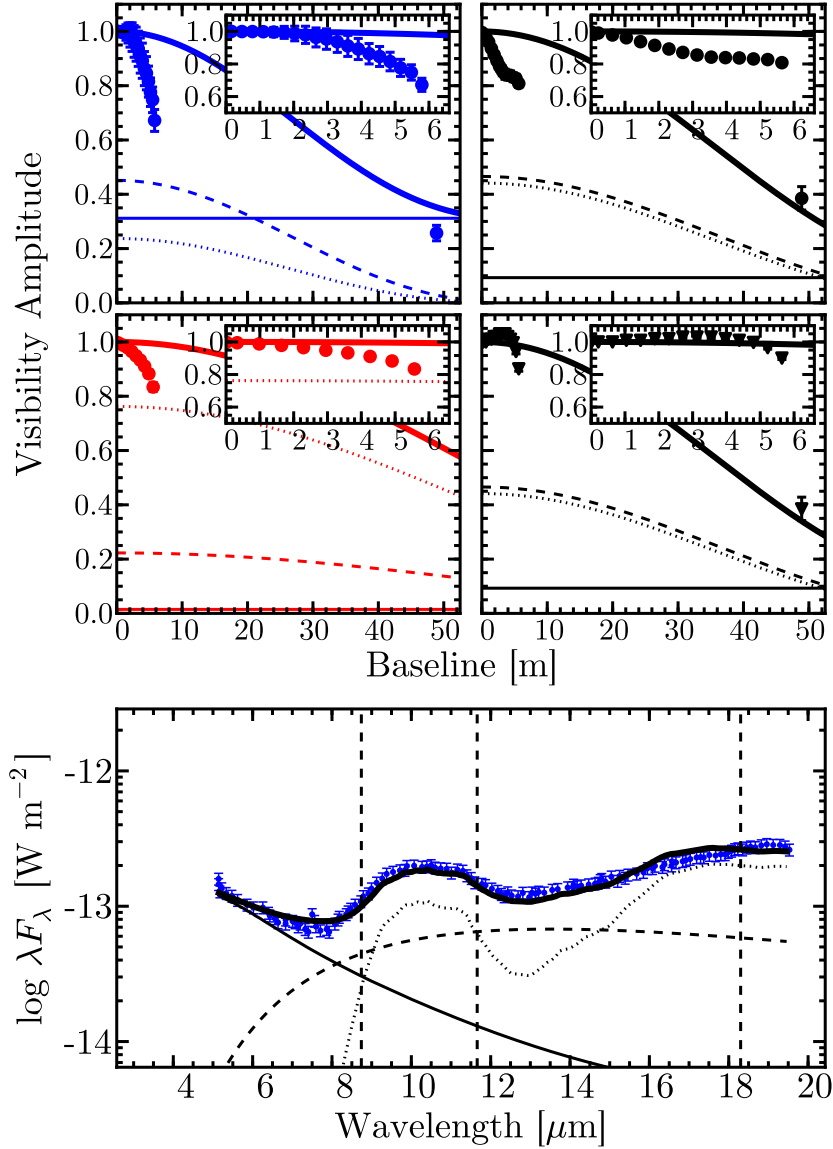}
    \caption{
The calculated SED and visibility curves for our realization of
\citet{Ratzka07}'s model. See Table~\ref{tab:Models} for model parameters.
The large bottom panel shows the SED, indicating the flux from the star
({\it thin solid line}\/), the optically thin disk wall atmosphere ({\it
dotted line}\/), the optically thick, directly illuminated  wall ({\it
dashed line}\/), and the total ({\it thick solid line}\/). The flux
density data are also shown ({\it blue circles}\/). In the top half of the
figure, we show our four separate observations: 8.74 ({\it blue, top
left}\/), 11.66a ({\it black, top right}\/), 11.66b ({\it black, bottom
right}\/), and 18.30 ({\it red, bottom left}\/). In each, the flux weighted
visibilities of the different model components are indicated by different
line types, as in the SED plot. The insets in the top right of each panel
show the data presented in this work, while the long baseline point at
$\sim$45~m is from \citet{Ratzka07}.  Vertical, dashed lines in the SED
panel show our wavelengths of observation.
    \label{fig:Ratzka}}
\end{figure}

We present the calculated SED and visibility curves for our realization of this
model in Figure~\ref{fig:Ratzka}. The large bottom panel shows the SED,
indicating the stellar flux density ({\it thin solid line}\/), the optically
thin, transitional disk wall atmosphere ({\it dotted-dashed line}\/), the
transitional disk wall ({\it dotted line}\/), and the total flux ({\it thick
solid line}\/). The flux density data are also shown ({\it blue circles}\/).
In the top half of the figure, we show our four separate observations: 8.74
({\it blue, top left}\/), 11.66a ({\it black, top right}\/), 11.66b ({\it
black, bottom right}\/), and 18.30 ({\it red, bottom left}\/). In each, the
flux weighted visibilities of the different model components are indicated by
different line types, as in the SED plot. The insets in the top right of each
panel show the data presented in this work, while the long baseline points at
$\sim$45~m are obtained from \citet{Ratzka07}.  Vertical, dashed lines in the
SED panel show our wavelengths of observation.

Our realization of \citet{Ratzka07}'s does not perfectly match the model SED
presented in \citet{Ratzka07} (particularly at $16-18$~\um), but does a
reasonably good job considering the complex disk properties (e.g., dust
settling) we did not include.  Our model visibilities do reproduce well those
presented in \citet{Ratzka07} at long baselines at 8.74~\um\ and 11.66~\um\
(see their Figure~8 for comparison). As can be seen in this figure, the
\citet{Ratzka07} data show the emission at $\sim$8.74~\um\ to be more resolved
than that at $\sim$11.66~\um, though the model of \citet{Ratzka07} does not
reproduce this behavior.  We see this trend of very-resolved 8.74~\um\ emission
in our own observations as well.

It is clear, however, that this small cavity model is inconsistent with nearly
all our data. Though it is consistent with one epoch of our 11.66~\um\ data
(1166b), it is very inconsistent with all the other observations. It is
unresolved at all baselines and all wavelengths relevant to this work. While we
did not include an outer disk, this does not impact the observed discrepancy
because the emission from these cool, outer regions is not relevant for the
wavelength ranges we are interested in. This omission may have slightly
worsened our reproduction of the small cavity model's 18.30~\um\ emission, but
outer disk components at cooler temperatures will not cause the model to show
an 8.74~\um\ component as resolved as our data show.

\begin{figure}
    \epsscale{0.7}
    \plotone{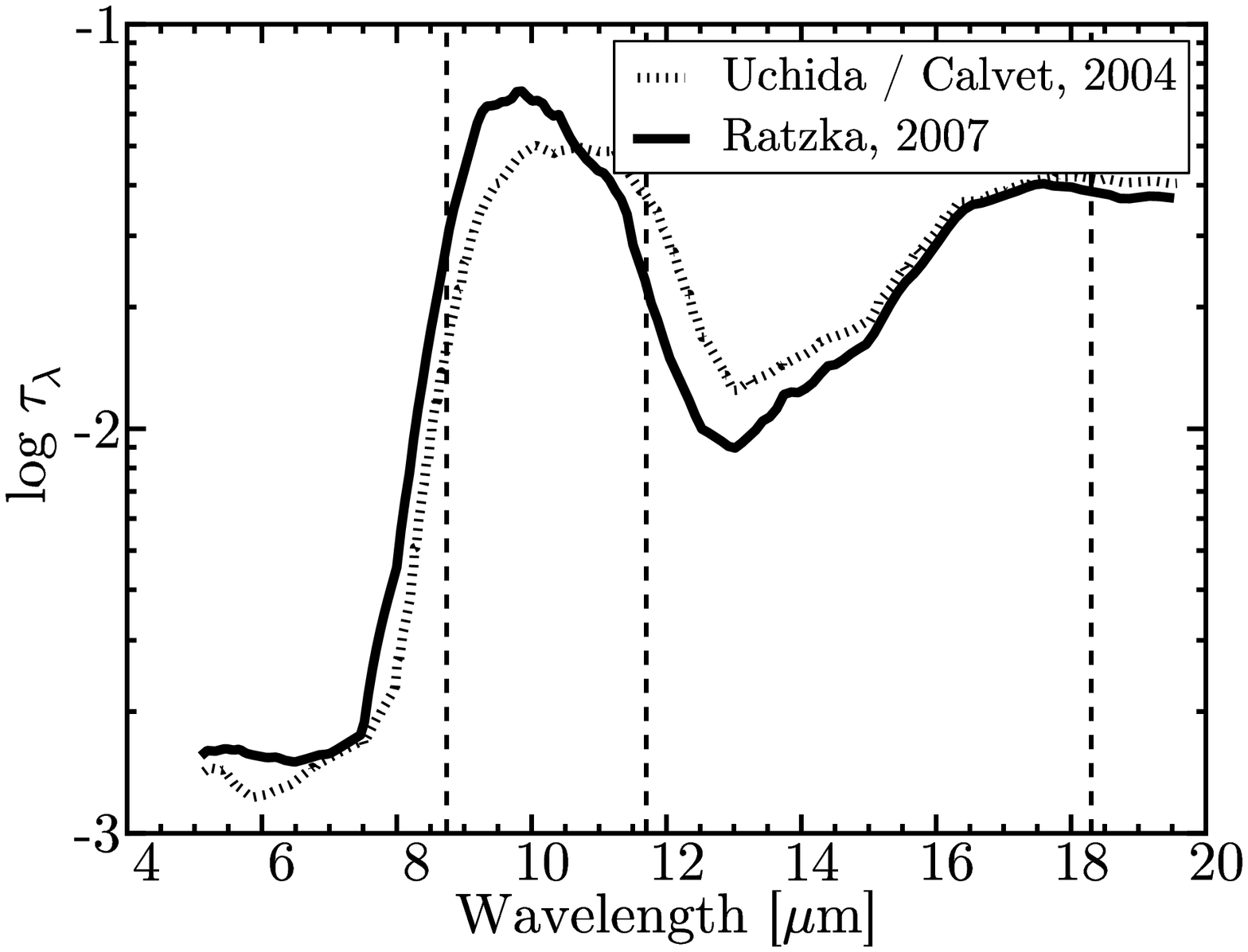}
    \caption{
The optical depth used by the works that initially presented large central
cavity \citep{Calvet02,Uchida04} and small central cavity \citep{Ratzka07}
models to explain the mid-IR emission from the \twhya\ disk system.  We
note the discrepancies in choice of optical depth by these authors. As the
optical depth and disk geometry are degenerate in model-fitting an
unresolved SED, these two geometrically distinct models are both able to
reproduce the SED\@. To examine the relative differences between choices of
optical depth, we have scaled the \citet{Ratzka07} optical depths so that
the mean values of each optical depth plot match. 
\label{fig:Opacity}}
\end{figure}

We suspect that the difference in dust opacity choices is the principal reason
that two distinct disk geometries---the $\sim$4~AU holes in \citet{Calvet02}
models, versus the $\lesssim$1~AU hole inferred by \citet{Ratzka07}---can
reproduce the same SED\@.  In Figure~\ref{fig:Opacity}, we compare the optical
depths used by \citet{Ratzka07} to those used by \citet{Calvet02,Uchida04}. In
order to examine the differences between the opacity choices in these two
works, we have converted the optical depths from \citet{Ratzka07} to a
dimensionless optical depth with an assumption of surface density (informed
from their work).  We then scaled this value so that the mean values of optical
depth for \citet{Calvet02} and \citet{Ratzka07} match, thus facilitating
comparison. The \citet{Ratzka07} optical depths show a relative paucity around
$11-12$~\um\ that explains why a smaller ($\lesssim$1~AU versus $\sim$4~AU for
a large cavity model) transitional disk rim is used in that model. We note that
the optical depths used by \citet{Calvet02,Uchida04} were derived
simultaneously with the other disk parameters, while those used by
\citet{Ratzka07} were not.

\subsection{New Disk Model}
\label{sec:NewDiskModel}

Neither of the previous models considered \citep{Calvet02,Ratzka07} adequately
fit our new observations. The hybrid model of \citet{Akeson11} also cannot
explain our very resolved 8.74~\um\ data. This is not surprising, as
\citet{Akeson11}'s model was designed to fit simultaneously the two model types
proffered by \citet{Calvet02,Ratzka07}. A new model is required.  In order to
produce more resolved emission at 8.74~\um, without producing heavily resolved
emission at larger wavelengths,  we need a hot component at large stellocentric
radius. We will show that this ``hot'' component must be considerably hotter
than predicted by equilibrium disk models. 

We adopt the optical depths from \citet{Uchida04} (see
Section~\ref{sec:ExistingModels}). We assume a Kurucz stellar atmosphere with
the scaling left as a free parameter. In addition to the components described
in Section~\ref{sec:ExistingModels}, we include an additional optically thick
ring ($R_\mathrm{hot}$ and $T_\mathrm{hot}$). We also vary the surface density
of the optically thin inner disk, $\Sigma_\mathrm{in}$, but leave the dust
sublimation radius $R_\mathrm{in}$ used by \citet{Calvet02,Uchida04} unchanged,
at \sublimationRadius~AU\@.  We allowed the following parameters to vary freely
(between physically reasonable limits) to generate our best-fit model: Kurucz
Scaling, $R_\mathrm{wall}$, $f_\mathrm{wall}$, $T_\mathrm{wall}$,
$\Sigma_\mathrm{in}$, $R_\mathrm{hot}$, $f_\mathrm{hot}$, and $T_\mathrm{hot}$.
For definitions of these parameters, see Table~\ref{tab:Models}.

\begin{figure}
    \epsscale{0.7}
    \plotone{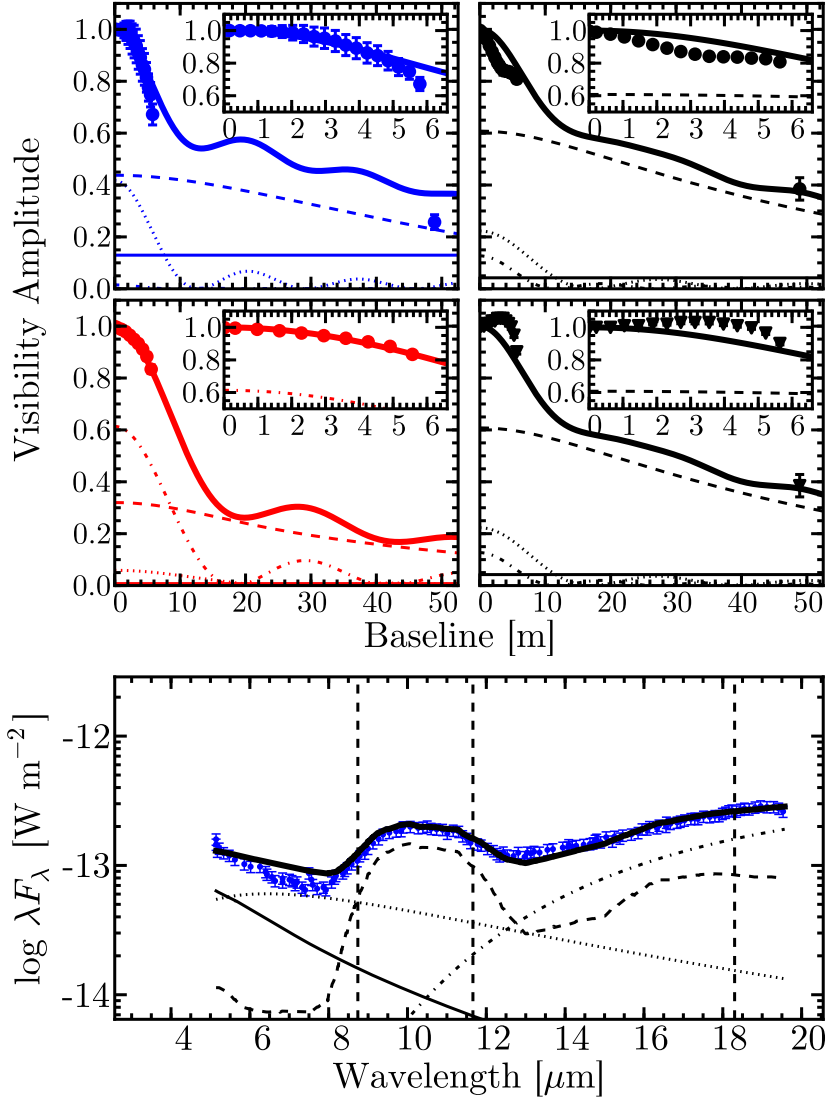}
    \caption{
      The calculated SED and visibility curves for the new model presented in
      this work.  See Table~\ref{tab:Models} for model parameters.  The large
      bottom panel shows the SED, indicating the flux from the star ({\it thin
      solid line}\/), the optically thin inner disk ({\it dotted line}\/), the
      optically thick, directly illuminated wall ({\it dashed line}\/), the
      hot, optically thick ring ({\it dashed-dotted line}\/), and the total
      ({\it thick solid line}\/). The flux density data are also shown ({\it
      blue circles}\/).  In the top half of the figure, we show our four
      separate observations: 8.74 ({\it blue, top left}\/), 11.66a ({\it black,
      top right}\/), 11.66b ({\it black, bottom right}\/), and 18.30 ({\it red,
      bottom left}\/). In each, the flux weighted visibilities of the different
      model components are indicated by different line types, as in the SED
      plot. The insets in the top right of each panel show the data presented
      in this work, while the long baseline point at $\sim$45~m is from
      \citet{Ratzka07}.  Vertical, dashed lines in the SED panel show our
      wavelengths of observation. 
\label{fig:Arnold}}
\end{figure}

After a thorough exploration of parameter space, we found a model that explains
the existing data. The observables associated with this model are presented in
Figure~\ref{fig:Arnold}. In this figure we show the calculated SED and
visibility curves for the model presented in this work.  This model is a large
cavity model, similar to \citet{Calvet02} and consistent with the results of
\citet{Hughes07}. The directly illuminated disk wall is represented by a ring
of optically thick emission at \arnoldWallRadius~AU, and, as in
\citet{Calvet02}, there is an optically thin disk of emission at low surface
densities that extends from the dust sublimation radius to the outer disk wall
($\sublimationRadius-\arnoldWallRadius$~AU). The crucial addition to this model
is an additional optically thick ring of emission inside the disk wall, at
$\sim$\hotRingRadius~AU\@. In order to fit our 8.74~\um\ data, this ring is at
a hotter than equilibrium temperature (which would be $\sim$150~K, compared to
$\sim$\hotRingTemperature~K for the hot ring).  We also find that the best fit
is achieved with an unscaled Kurucz stellar model, unlike the scalings of
$\sim$2 required to fit the data using the models of
\citet{Calvet02,Uchida04,Ratzka07}.

We note that with this model we produce a good fit to the SED, discrepant only
in the 6-8~\um\ region where possible stellar variability is a larger concern.
Even so, the discrepancy in the SED fit is comparable in magnitude to that of
the \citet{Uchida04} deviation from observations in the $\sim$13~\um\ region.
We note that our fit to the long baseline mid-IR visibility amplitudes from
\citet{Ratzka07} is comparable to the model visibilities produced by
\citet{Ratzka07}'s own model. Most notably, however, we point out the
simultaneous high quality fit of the long baseline mid-IR visibility amplitudes
with the short baseline mid-IR visibility amplitudes presented in this work.
Again, this was accomplished by forcing the ring to be hot enough to peak at
shorter wavelengths---closer to 8.74~\um---allowing the short baseline
8.74~\um\ emission to be resolved, as we see in both our data and the VLTI data
of \citet{Ratzka07}. Additionally, despite the increased degrees of freedom,
the reduced chi squared value for our new disk model is $\sim$1\% better than
the value we calculate for the Calvet model. The reduced chi squared value for
the Ratzka model is considerably larger. The value of the chi squared is
largely driven by the unresolved epoch of our 11.66~\um\ micron data, however.
The less-resolved Calvet model fits this epoch slightly better. If we fit the
average of the two 11 micron epochs, instead of attempting to fit both
simultaneously, our new model has a reduced chi squared that is a factor of ~3
lower.

\subsection{Disk + Companion Model}
\label{sec:DiskCompanionModel}

The models described thus far are all symmetric about the azimuth. As described
in Section~\ref{sec:Calibration}, the reason we made this assumption was to
decrease the uncertainty associated with our measurements of baseline-dependent
visibility and to make our results more straight-forward to interpret. However,
our solution of a hotter-than-equilibrium, geometrically thin ring of emission
at a radius just inside the directly illuminated disk wall does not seem
physical.  That is, we cannot think of a physical process that could heat an
annulus of matter far beyond the temperature achieved for a disk in equilibrium
with stellar emission.

A self-luminous companion, on the other hand, can lead to such heating.  Such
an object would clearly not be azimuthally symmetric.  The presence of a
companion should thus manifest itself in our data through departures from
azimuthal symmetry as a function of wavelength.

If all of our model components except the companion are azimuthally symmetric,
we can predict relative degrees of asymmetry in the different wavelengths of
observation. We expect the azimuthal asymmetry to
be largest at 8.74~\um; at 11.66 and 18.30~\um, the flux contribution from the
hot inner component is small compared to fluxes from other, symmetric model
components (see Figure~\ref{fig:Arnold}).

To test this, we create a model with a companion that has the same temperature
and surface area as our added ring component and thus the disk plus companion
model produces an SED identical to the model described in
Section~\ref{sec:NewDiskModel}. For this reason we can utilize only our
visibility data to constrain properties of the companion's stellocentric radius
and position angle (PA); the SED will not change as these parameters are
varied.

We then calculate the degree of asymmetry generated by the presence of a
companion in this model. Similarly, we measure the asymmetry in our reduced
data. We follow the procedures outlined in Section~\ref{sec:Reduction} with one
exception. Instead of performing an azimuthal average of the power image, we
calculate the radial average as before, but include only small regions (of
width 45 degrees) of azimuth separated by 23 degrees. That is, we calculate the
radial average (power as a function of baseline) in different ``wedges'' of
azimuth.  We perform these steps for both a synthetic image of the disk plus
companion model, and for our data.

Using the azimuth wedges for both synthetic model and reduced data, we compare
the disk plus companion model to the data. We fix all the components of the
model except the companion stellocentric radius and position angle.  We then
perform a best-fit chi squared minimzation over these two freely varying
parameters (PA and stellocentric radius) for all epochs of our data (see
Table~\ref{tab:Observations}). We find a best-fit companion radius of
$\sim$\hotCompanionRadius~AU, and a PA of $\sim$90~degrees. We note that as we
have limited phase information using this technique, that we are unable to
distinguish between PAs separated by 180~degrees. Thus, our best-fit PA is
$\sim$90~degrees or $\sim$270~degrees. Position angle is measured
counter-clockwise eastwards from north.

Since our data span two epochs separated by $\sim$720~days, we do not fit
either epoch of the 11.66~\um\ data perfectly. Further, if a companion indeed
explains our data, then this object would undergo orbital evolution between our
two epochs. For these reasons, we additionally fit PA separately for each epoch
for a fixed radius: \hotCompanionRadius~AU, the best-fit companion radius for
all the data. This exercise yields a best-fit PA of 90~degrees for our 2007
data (874 and 1166a), but the PA is not strongly constrained in the 2009 data
(1166b and 1830). This is unsurprising, as the 2009 epoch does not contain
short-wavelength observations in which the flux from the hot companion would be
significant.

\begin{figure}
    \epsscale{0.7}
    \plotone{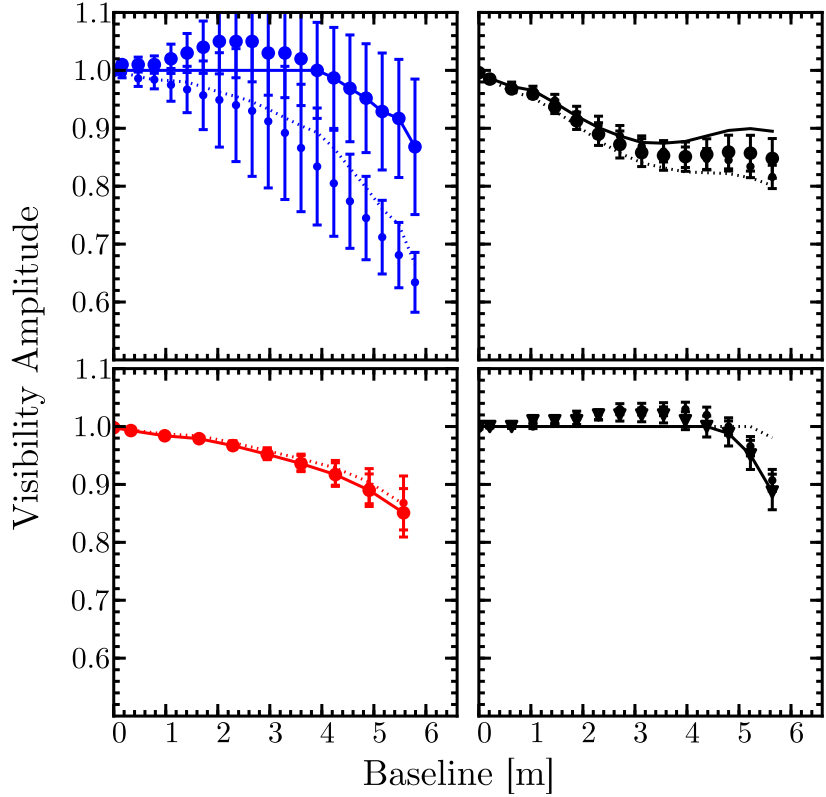}
    \caption{
This figure shows the asymmetry in our observations.  The two sets of
points with error bars in the figure show the normalized, calibrated
visibility amplitude as a function of baseline, as in
Figures~\ref{fig:Calvet},~\ref{fig:Ratzka},~and~\ref{fig:Arnold}. Here,
though, instead of azimuthally averaging as described in
Section~\ref{sec:Reduction}, we perform a radial average along two narrow
slices of angle in azimuth, offset by 90 degrees. Each curve thus shows
how resolved the image is along each direction in the Fourier power image.
This figure shows that the 8.74~\um\ emission is very resolved along one
direction, and quite unresolved along another orthogonal direction. These
data are inconsistent with an azimuthally symmetric ring of emission. The
solid and dotted lines show the ``de-trended'' (see
Section~\ref{sec:DiskCompanionModel}) predicted degree of asymmetry from
a best-fit disk plus companion model.
\label{fig:AsymDetrended}}
\end{figure}

To more clearly illustrate asymmetry in our multi-epoch data and to ameliorate
apparent discrepancies due to fitting our disk model to the combined 2007 and
2009 datasets, we perform a simple ``de-trending'' of the model visibility
amplitudes.  That is, we force the synthesized visibilities to match the shape
of the visibilities obtained through our data reduction by applying a
multiplicative scaling at each baseline. As visibilities greater than one are
unphysical, we set detrended model values that exceed one to be equal to one.
In Figure~\ref{fig:AsymDetrended}, we show the expected magnitude of the
asymmetry and compare this to the magnitude of the asymmetry in our
observations. 

In this figure, the points with error bars show the degree of asymmetry
revealed in our data reduction, using the wedge comparison technique described
above. We choose the best-fit PA for each epoch for one wedge, and the best-fit
PA plus 90 degrees for the second wedge.  Though not highly statistically
significant, this figure shows that the asymmetries are consistent with our
expectation from an asymmetric source of emission at short wavelengths: that
the asymmetry in the 8.74~\um\ data is greatest, smaller at 11.66~\um, and yet
smaller at 18.30~\um. The solid and dotted lines are the visibilities obtained
by generating a synthetic companion model and directly extracting the
normalized visibilities via Fourier techniques, described in the preceding
paragraphs.

\begin{figure}
    \epsscale{0.7}
    \plotone{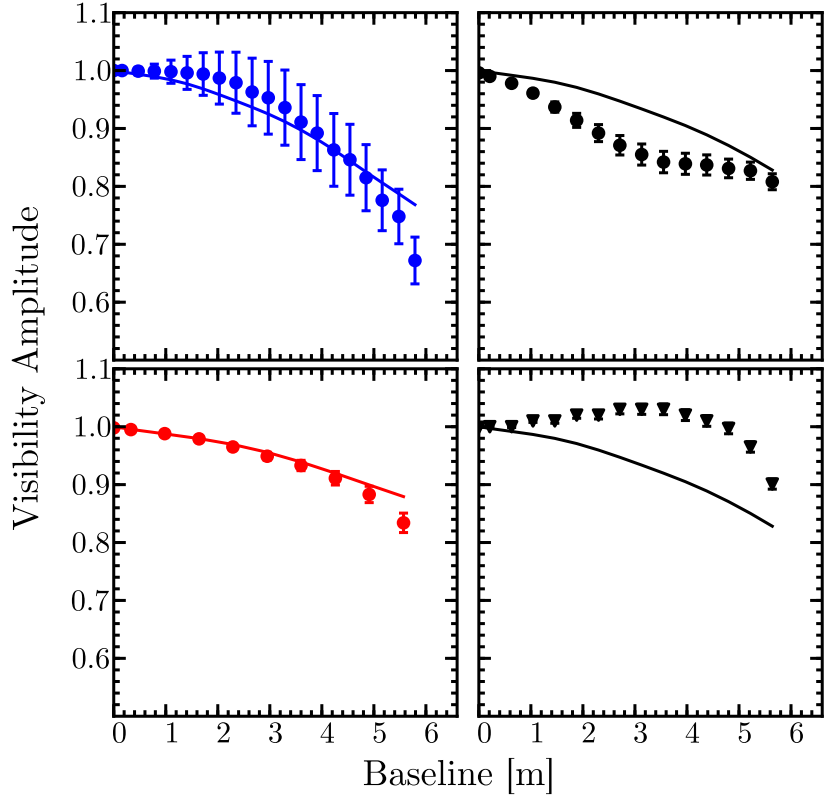}
    \caption{
This figure shows the synthetic visibilities generated by our disk plus
companion model, compared to our data.  While we fitted models to data in
azimuthal wedges, we plot the azimuthally averaged data for simplicity.
The points with error bars in the figure show the normalized, calibrated
visibility amplitude as a function of baseline, as in
Figures~\ref{fig:Calvet},~\ref{fig:Ratzka},~and~\ref{fig:Arnold}. As in
the previous figures, we show our four separate observations: 8.74 ({\it
blue, top left}\/), 11.66a ({\it black, top right}\/), 11.66b ({\it black,
bottom right}\/), and 18.30 ({\it red, bottom left}\/). Our data are indicated
by unconnected points, and the synthetic model visibilities are shown with
solid lines. Note that we have not accounted for potential disk
variability between the two observed epochs, and hence our model tends to
fit the average of the two 11.66~\um\ datasets.
\label{fig:VisibCompanion}}
\end{figure}

\begin{figure}
    \epsscale{0.7}
    \plotone{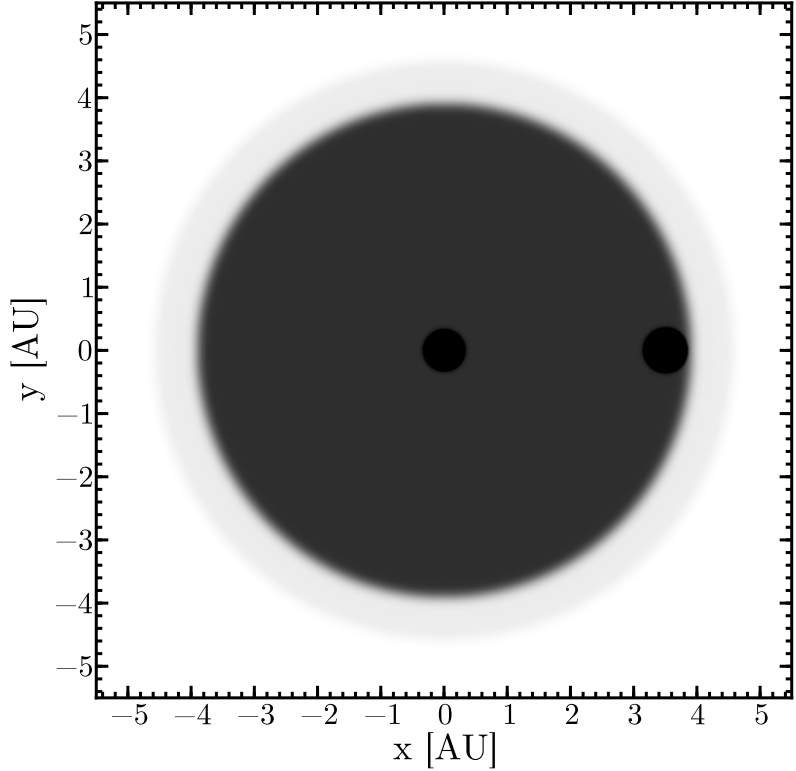}
    \caption{
A synthesized image of our disk + companion model at 8.74~\um. For model
parameters, see Table~\ref{tab:Models}. Most of the model components can be
clearly seen; only the completely cleared region inside the dust sublimation
radius ($\sim$\sublimationRadius~AU) is absent at the displayed resolution.
Several model components are visible: the central star; the optically thin disk
extending from the dust sublimation radius to the transition disk wall at
\arnoldWallRadius~AU; the directly illuminated, optically thick wall at
\arnoldWallRadius~AU (faint, but visible, at 8.74~\um), represented as a single
temperature ring; and the point source companion, producing optically thick hot
emission at \hotCompanionRadius~AU, required to explain our very resolved
8.74~\um\ observations. The companion has the same surface area as the hot ring
model described in Section~\ref{sec:NewDiskModel} and Table~\ref{tab:Models}.
This panel shows our model at 8.74~\um; the best-fit position angle of the
companion is shown for this epoch. In order to make the geometrically small
elements of the model visible in this illustration, we have convolved the model
with a PSF with full width half maximum (FWHM) $\sim$10~mas.
\label{fig:SynthImage}}
\end{figure}

We present the azimuthally averaged visibilities from our disk plus companion
model with best-fit values for companion stellocentric radius and PA in
Figure~\ref{fig:VisibCompanion}, for comparison with our traditional disk
models (Figures~\ref{fig:Calvet},~\ref{fig:Ratzka}, and~\ref{fig:Arnold}). We
present a synthetic image of this model in Figure~\ref{fig:SynthImage}.

We also examined the closure phases of our data. Closure phases---a measure of
the complex phase that removes the phase ambiguity produced by the turbulent
atmosphere---are used to discover and constrain asymmetric structure in
observed objects (e.g., a companion).  We found that the closure phases of our
data are insensitive to structures on the small scales we are examining. The
closure phases are sufficiently noisy to be largely insensitive to asymmetric
structure more compact than $\sim$100~mas (our putative companion lies at
$\sim$\ourPredictedSeparation~mas).

\begin{deluxetable}{clrrrrr} 
\tabletypesize{\scriptsize}
\tablecolumns{7} 
\tablewidth{0pc} 
\tablecaption{Model Properties} 
\tablehead{ 
    \colhead{}      & 
    \colhead{}      & 
    \colhead{}      & 
    \colhead{Calvet / Uchida} & 
    \colhead{Ratzka} &
    \colhead{New Disk Model}  &
    \colhead{Disk + Companion Model}   }
\startdata 
 (1) &  Kurucz Scaling           &            &  \calvetKuruczScale      & \ratzkaKuruczScale     & \arnoldKuruczScale       & \arnoldKuruczScale      \\
 (2) &  $R_\mathrm{in}^*$        & (AU)       &  \sublimationRadius      &  -                     & \sublimationRadius       & \sublimationRadius      \\
 (3) &  $T_\mathrm{in}^*$        & (K)        &  \sublimationTemperature &  -                     & \sublimationTemperature  & \sublimationTemperature \\
 (4) &  $\Sigma_\mathrm{in}$ /
  $\Sigma_\mathrm{in,Calvet}^*$  &            &   1                      &  -                     & \arnoldSigma             & \arnoldSigma            \\
 (5) &  $R_\mathrm{wall}$        & (AU)       &  \calvetWallRadius       & \ratzkaWallRadius      & \arnoldWallRadius        & \arnoldWallRadius       \\
 (6) &  $f_\mathrm{wall}$        &            &  \calvetfRatio           & \ratzkaWallfRatio      & \arnoldfRatio            & \calvetfRatio           \\
 (7) &  $T_\mathrm{wall}$        & (K)        &  \calvetWallTemperature  & \ratzkaWallTemperature & \arnoldWallTemperature   & \arnoldWallTemperature  \\
 (8) &  $R_\mathrm{atm}^*$       & (AU)       &   -                      & \ratzkaAtmRadius       &  -                       &  -                      \\
 (9) &  $f_\mathrm{atm}^*$       &            &   -                      & \calvetfRatio          &  -                       &  -                      \\
(10) &  $T_\mathrm{atm}^*$       & (K)        &   -                      & \ratzkaAtmTemperature  &  -                       &  -                      \\
(11) &  $\Sigma_\mathrm{atm}$ /
  $\Sigma_\mathrm{in,Calvet}^*$  &            &   -                      & \ratzkaAtmSigma        &  -                       &  -                      \\
(12) &  $R_\mathrm{hot}$         & (AU)       &   -                      &  -                     & \hotRingRadius           & \hotCompanionRadius     \\
(13) &  $f_\mathrm{hot}$         &            &   -                      &  -                     & \hotRingfRatio           &  -                      \\
(14) &  $T_\mathrm{hot}$         & (K)        &   -                      &  -                     & \hotRingTemperature      & \hotRingTemperature     \\
(15) &   Opacities               &            &   C                      &  R                     &  C                       &  C                      \\
(16) &   Position Angle (2007)   & ($^\circ$) &   -                      &  -                     &  -                       &  90                     \\
(17) &   Position Angle (2009)   & ($^\circ$) &   -                      &  -                     &  -                       &  -                      \\
\enddata 

\tablecomments{
The models used to generate SEDs and visibilities
presented in this work. Columns separate models that are ``Calvet-like'',
``Ratzka-like'', or newly presented in this work. Rows describe the different
components in the model, and are (a $^*$ indicates the component is optically
thin, otherwise it is optically thick):
(1) the scaling factor applied to the Kurucz model, described in Section~\ref{sec:Modeling};
(2) the dust sublimation radius, i.e., the inner radius of the optically
thin disk; 
(3) the temperature at the dust sublimation radius;
(4) the surface density of the inner, optically thin disk, compared to the
surface density used by \citet{Calvet02};
(5) the radius of the directly illuminated optically thick wall;
(6) the ratio of width to radius for the transition disk wall;
(7) temperature of the optically thick wall;
(8) radius of the optically thin wall atmosphere;
(9) the ratio of width to radius for optically thin wall atmosphere;
(10) temperature of the optically thin wall atmosphere;
(11) the surface density of the optically thin wall atmosphere, compared
to the surface density used by \citet{Calvet02};
(12) radius of the optically thick, hotter-than-equilibrium ring inside the
transition disk wall. This component is only used for the model
presented in this work;
(13) the ratio of width to radius for the optically thick,
hotter-than-equilibrium ring;
(14) temperature of the optically thick, hotter-than-equilibrium ring;
(15) opacities used in each model---``C'' for Calvet, ``R'' for Ratzka.
See Section~\ref{sec:SmallCavityModels} 
(16) The best-fit position angle of our companion, for our 2007 data
(17) The best-fit position angle of our companion, for our 2009 data
}
\label{tab:Models}
\end{deluxetable} 

\section{Discussion}
\label{sec:Discussion}

The presence of a companion in the \twhya\ system is consistent with theories
of transition disk dissipation
\citep[e.g.,][]{GoldreichTremaine82,Bryden99,Rice03}.  The fitted stellocentric
radius of our putative companion is $\sim$\hotCompanionRadius~AU, inside the
optically thick disk wall at $\sim$\arnoldWallRadius~AU\@. The Keplerian period
for such a circular orbit would be $\sim$\hotCompanionOrbitalPeriod~days,
assuming a mass for \twhya\ of $0.7$~$M_\odot$ \citep{Webb99}. The time between
our two epochs is $\sim$720~days, and so during this time the change of the PA
of the companion would be $\sim$\deltaThetaInBetweenEpochs~degrees. Since we do
not have an estimate of PA for our second epoch, we cannot test for this
orbital evolution in our data.  Though the 2009 data do not constrain a value
for PA, the data are consistent with orbital motion of a companion.  This is
because the companion's flux contribution at 11.66~\um\ and 18.30~\um\ is less
significant; another epoch of 8.74~\um\ observations would enable this
exercise.

We have not quoted uncertainties on any of our fitted parameters. Our
parameterizations for the reproduction of the \citet{Ratzka07} model and
our new disk model include 7 and 9 freely varying parameters,
respectively.  With parameter spaces of increasing dimensionality, model
uniqueness and degeneracy of parameters becomes increasingly problematic.
High dimensional chi-squared surfaces are complex, and may contain many
local minima. To obtain our best-fit parameters, we employ grid-based
chi-squared minimizations. We obtain a best-fit PA and stellocentric
radius of the companion by simply substituting the hot ring in our new
disk model ($\S$~\ref{sec:NewDiskModel}) for the emission from an
unresolved source.  In doing this, we fix all of the many free parameters
in our model (radius of optically thick disk rim, temperature of model
components, etc.).  Obtaining the uncertainties for PA and radius from
a chi-squared surface assuming only one or two varying parameters instead
of the actual, larger number, would not provide an accurate estimate of
the error in these parameters. Since modeling the multi-dimensional
chi-squared surface is computationally prohibitive using our methods, we
do not report formal confidence intervals on our model parameters.

While the flux from the outer regions of the disk (at 11.66~\um\ and longer
wavelengths) is rather sensitive to changes in $R_\mathrm{wall}$ and
$T_\mathrm{wall}$ (and has been modeled previously, \citet[][e.g.]{Calvet02}),
the quality of the fit to the data depend less strongly on $T_\mathrm{comp}$
and $R_\mathrm{comp}$, due in part to the variable nature of \twhya\ at the
shorter wavelengths. We find that while changes of just $\sim$10s of Kelvins or
$\sim$tenths of an AU in the disk cavity wall can lead to an unacceptable fit,
changes of a similar magnitude in the hot disk or companion cause a much
smaller effect. In fact, a temperature for the hot companion as low as 430~K
produces a lower-quality but still acceptable fit to our data: a deficit of
flux at wavelengths shorter than 6-8~\um\ can be explained by a larger stellar
contribution. Though some model parameters are degenerate, we use this range of
temperatures (430-500~K) to investigate potential observables predicted by
planetary models.

Even if we assume our lowest acceptable value for $T_\mathrm{comp}$, the fitted
temperature of our companion is hotter than the disk equilibrium temperature at
its stellocentric radius ($\sim$\companionTempAtEquilibrium~K). Given our
fitted surface area of emission, and the assumption that this object emits like
a blackbody, the luminosity of this object is
$\sim$\luminosityCompanion~$L_\odot$. The age of the \twhya\ system is
$\sim$10~Myr \citep{Webb99}, which puts an upper limit on the age of the
putative companion.

These properties of our proposed companion are consistent with models of
planetary formation. The temperature of our companion (of age $\lesssim10$~Myr)
is consistent with the predicted temperatures from planetary thermal evolution
models presented in \citet{Fortney08} for a planet beginning from the core
accretion formation models of \citet{Hubickyj05}. The luminosity, however, is
sufficiently large to require ``hot start'' models with arbitrary initial
conditions; in this case, our companion's luminosity is consistent a mass of
8-10~\Mjup\ \citep{Fortney08}. Indeed, the work of \citet{MarleyFortney07}
suggests that such a large luminosity is only possible with hot-start models,
or if the companion is undergoing an accretion shock phase during the
core-accretion formation process. \citet{MarleyFortney07} state, however, that
luminosities for the hot-start planetary models are highly uncertain and depend
strongly on the initial entropy of the evolution tracks; this makes mass
determination of very young planets difficult. We also note that the surface
flux density of our companion as a function of wavelength is consistent with
those of a 500-700~K model presented in \citet{Fortney08}, except for effects
due to opacity, which we do not include (see their Figure~3).

The luminosity of our planet is feasible if it is undergoing an accretion
shock phase, though the magnitude of an object undergoing this phase is
highly uncertain, and this phase is short-lived \citep{MarleyFortney07}.
Another possible explanation for the high magnitude of the companion
luminosity is that the companion is surrounded by and accreting
circum-planetary material; an increase in the surface area of the emitting
region would produce a larger luminosity. 

Additionally, we have detected significant variability at 11.66~\um\ in our
multi-epoch observations.  The origin of this variability is unclear.  The
fitted size (using a ring emitting at a single wavelength) for the more
resolved epoch is $2.56\pm0.23$~AU, compared to the less resolved size of
$0.71\pm0.50$~AU\@.  At 11.66~\um, emission is dominated by the optically thin
transition disk cavity, at relatively small stellocentric radii; indeed, much
of the emission at this wavelength comes from the hot inner edge of the
optically thin disk, at the dust sublimation radius.  The variability is thus
likely due to changes in the properties of this inner region.
\citet{Muzerolle09} report ``remarkable mid-IR variability'' in the transition
disk LRLL~31, in the same wavelength range as our observed variable sized
emission and on timescales as short as one week.  \citet{FlahertyMuzerolle10}
propose models to explain this sort of mid-IR variability, using
non-axisymmetric perturbations in the disk (e.g., a warp with variable scale
height, or spiral wave).  \citet{Flaherty11} expands on the earlier work of
\citet{Muzerolle09} with extensive observations of LRLL~31 over many epochs in
the infrared. They find that the dust destruction radius stays relatively
constant, that accretion and IR excess vary over timescales of $\sim$weeks, and
that changes in scale height and/or warping of the inner disk is likely
responsible for the observed variation. They rule out accretion and stellar
winds as a cause for the disk changes, and conclude that the most likely
explanation for their observations is a companion or a dynamic interface
between the stellar magnetic field and the disk. In particular, they describe
how a companion, orbiting at an inclination relative to the disk, can drag dust
from the disk midplane in a periodic fashion, producing variable IR emission.
Given that the inner disk contributes most of the flux at 11.66~\um\ in our
model, variability in the disk itself---rather than variable flux from a
companion---is the most likely explanation, as in \citet{Flaherty11}. As we
only have one epoch of 8.74~\um\ data, we cannot determine if the variability
we observe persists at shorter wavelengths, though \twhya\ has also been
observed to be variable in the near-IR and optical as well \citep[][and
references therein]{Eisner10}.  A multi-epoch, multi-wavelength study of
\twhya, as in \citet{Flaherty11} for LRLL~31, would yield a greater
understanding of the physical mechanisms responsible for this variability.

Recently, other companions have been detected or inferred around transition
disk objects.  \citet{Huelamo11} presented evidence of a companion around
T~Chamaeleontis at $\Delta L=5.1$~mag and at a separation of 62$\pm$7 mas using
sparse aperture masking and adaptive optics at the VLT. \citet{Eisner09}
present spatially resolved mid-IR observations from \trecs\ of SR~21, at a
separation $\lesssim100$~mas.  They predict that in the $K$ band the companion
has $\sim5\%$ of the flux of the central star, and $\sim25\%$ at $L$.
\citet{KrausIreland11} detect a companion around LkCa~15 using non-redundant
aperture masking interferometry at the Keck-II telescope.  This object has
projected separations of 72, 101, and 88~mas at three different epochs, with a
contrast in the $L$ band of $\Delta L=4.7$~mag. \citet{KrausIreland11}'s
detection of a companion around LkCa~15 shows evidence of asymmetry and
variability in stellocentric radius and flux (though they state that some of
the asymmetrical smearing is indicative of poor data quality).  They suggest
that circumplanetary material is expected around the detected companion, as
both planet and star are accreting protoplanetary disk mass.

We predict that in the $L$ band the companion around \twhya\ will have
$\sim$\ourPredictedFluxFractionL\% of the flux of the central star ($\Delta
L\sim$\ourPredictedMagContrastL~mag), and in the $M$ band (4.7~\um) the
companion will have $\sim$\ourPredictedFluxFractionM\% of the flux. Our
companion would lie at a separation of $\sim$\ourPredictedSeparation~mas from
the central star.  The contrast would be even more favorable at longer
wavelengths; the hot companion's contribution to the SED peaks at around 7~\um.
Due to its proximity then, \twhya\ offers the possibility to detect a companion
at a smaller stellocentric radius than, but at comparable angular separation
to, these recent companion discovery publications.   High resolution mid-IR
observations (perhaps speckle interferometry or non-redundant aperture
interferometry at the Large Binocular Telescope) should verify our predictions.

We note that \citet{Evans12} have placed a lower limit on the $L'$-band
contrast of a companion in a separation range of 40-80~mas around \twhya\ at
5.28~mag, consistent with but on the faint end of our L-band contrast estimate
of $\sim$\ourPredictedMagContrastL~mag. There are several reasons why our
prediction for companion contrast at $L$ band could be too large, however. For
example, if some of the emission from the companion is optically thin, the
companion flux at $L$ band would be smaller (while maintaining the same flux at
$8$~\um), leading to a larger value for the contrast at $L$ while preserving
the quality of our fit to the SED\@. Some optically thin emission would be
evidence that this emission is partially due to circumplanetary matter, which
would be more spatially extended to produce the same flux as a blackbody
component.

\section{Conclusions}

We present new mid-IR, spatially resolved measurements of the transition
disk object \twhya, taken in a novel observing mode at the Gemini
telescope using the \trecs\ instrument. We observed this object using
speckle imaging and reduced the data using Fourier image analysis
techniques. Our individual exposures were short enough to freeze the
atmosphere's turbulence, allowing for sub seeing-limited observations. In
fact, due to high precision calibration of the PSF using our Fourier
methods, we probe spatial scales at the diffraction limit of the Gemini
telescope. At all our observational wavelengths, we resolve the science
target, \twhya. 

We recreate and present simple models of \twhya's disk from the literature.  We
analyze the compatibility of these models with our new data, and show that
existing models do not reproduce our very resolved 8.74~\um\ emission. We
create a new model that satisfactorily explains all the available data: our new
resolved mid-IR measurements, long baseline interferometric measurements at
$7$~mm \citep{Hughes07}, the flux density in the mid-IR, and long baseline
mid-IR visibility amplitudes presented by \citet{Ratzka07}. This model has a
large, relatively empty cavity as shown in previous works
\citep{Calvet02,Uchida04,Hughes07}, but also includes a self luminous companion
at stellocentric radius of $\sim$\hotCompanionRadius~AU to explain our highly
resolved 8.74~\um\ data. We note also that the behavior of our data is similar
to that of \citet{Ratzka07}'s (8.74~\um\ emission more resolved than 11.66~\um\
emission), who obtain mid-IR observations at another epoch (2005) and at longer
baselines ($\sim$45~m). 

In summary, we present new observations with 8.74~\um\ emission more resolved
than 11.66~\um\ emission at baselines of $\sim$6~m. This unexpected result is
consistent with other findings: \citet{Ratzka07} show $\sim$8~\um\ emission
more resolved than $\sim$12~\um\ emission at much longer baselines. If a
companion is responsible for this emission, observations should show an
asymmetry, most significant at the peak emission wavelengths of the companion.
Our data are consistent with this expectation of asymmetry, but we cannot
constrain the details of this asymmetry well. We provide estimates for
separation and flux ratio of a putative companion, but these are uncertain due
to large number of components in our model, and because the companion is not
well-resolved.  Evidence for a luminous source at a large stellocentric radius
is fairly strong, but our constraints on its properties are weaker.

We measure the size of the emission in our observations at each wavelength, and
detect temporal variability at the $\gtrsim$2.5-$\sigma$ level between two
epochs at 11.66~\um. We speculate that this variability reflects variable
accretion through the inner disk, signatures of dynamical perturbations, and /
or perhaps a variable brightness of a self-luminous companion.

The authors would like to thank Laird Close, Phil Hinz, George Rieke, and the
referee for helpful suggestions and critiques.  This work is based on
observations obtained at the Gemini Observatory.  TJA was supported by the
National Science Foundation through a Graduate Research Fellowship. JAE
acknowledges support from an Alfred P. Sloan Research Fellowship. Our Gemini
Program IDs are GS-2008A-Q-18 and GS-2007A-Q-38.

\end{document}